\documentclass[11pt,a4paper]{article}

\usepackage[T1]{fontenc}
\usepackage[utf8]{inputenc}
\usepackage[english]{babel}
\usepackage{graphicx}%
\usepackage{multirow}%
\usepackage{amsmath,amssymb,amsfonts}%
\usepackage{amsthm}%
\usepackage{mathrsfs}%
\usepackage[title]{appendix}%
\usepackage{xcolor}%
\usepackage{textcomp}%
\usepackage{manyfoot}%
\usepackage{booktabs}%
\usepackage{algorithm}%
\usepackage{algorithmicx}%
\usepackage{algpseudocode}%
\usepackage{listings}%
\usepackage{hyperref}%
\usepackage{geometry}
\usepackage{amsmath,amssymb,amsthm}
\usepackage{mathtools}

\usepackage{graphicx}
\usepackage{caption}
\usepackage{subcaption}
\usepackage{booktabs}

\usepackage[numbers,sort&compress]{natbib}

\title{The effects of boundary conditions on Rindler's spectral anomaly}

\author{
  M. A. Estévez$^{1, 2}$, and E. Sadurní$^1$ \\
  \small $^{1}$Instituto de Física, Benemérita Universidad Autónoma de Puebla,\\ \small Apartado Postal J-48, 72570 Puebla, México \\
  \small $^{2}$Facultad de Ciencias Físico Matemáticas, Benemérita Universidad Autónoma de Puebla,\\ \small 72570 Puebla, México \\
  \small \texttt{sadurni.emerson@gmail.com}
}
\date{December, 2025}

\begin{document}

\maketitle

\begin{abstract}
Rindler's metric is an interesting way to incorporate a set of uniformly accelerated observers into space-time coordinates; this is consistent with special and general relativity. It is known that such an acceleration gives rise to the famous Unruh effect. Interestingly, its Galilean limit already shows the appearance of quantized modes for particles in free space, given by Airy functions. This happens when a wall or boundary condition is moving in an accelerated trajectory in free space and in the presence of a field. Here we show that such a boundary, when viewed as a material obstacle in motion, gives rise to quantized modes for the Klein-Gordon and Maxwell fields, as long as the boundary does not touch the singularity at the Rindler wedge. This corresponds to a quantum-mechanical problem with an anomalous fall-to-the-origin potential $-1/x^2$ supplemented with a Dirichlet condition. We provide further mathematical analysis regarding the completeness of the solutions in terms of Hankel functions $H^{(1)}$ of imaginary index and argument, and clarify the nature of the corresponding Sobolev spaces when the boundary condition disappears for the accelerated observer. A detailed interpretation of the transition amplitudes is given in connection with particle production obtained from a Bogoliubov transformation.
\end{abstract}
\section{Introduction}

The Unruh effect predicts how uniformly accelerated observers will perceive a change in the vacuum state of a field by means of observables such as the number operator. The effect was originally studied for scalar fields and recurrently reviewed by particle physicists in \cite{Unruh,Crispino}, but its realization is yet to be confirmed experimentally. By examining the particle production mechanism given by the Bogoliubov transformation, we can conclude that the result should be similar for other types of quantized fields, including Maxwell's theory. Therefore, an accelerated polarizer interacting with the inertial vacuum state should experience an effective radiation field and modify its motion as a reaction to the radiation pressure.

The Unruh effect has motivated experimental \cite{Lynch,Lima,Jiazhong} and theoretical \cite{Gooding,MartinMartinez,Cozzella,Carballo-Rubio} investigations on field quantization in curved spaces, as well as important discussions on its plausible detection \cite{Ford}. We are interested in studying the mathematical implications of the Unruh effect for photons and Klein-Gordon particles, interacting with obstacles, mirrors and polarizers. Our focus is to resolve the problem of quantized bound states for the anomalous $-1/x^2$ potential and investigate the mathematical implications associated with this scenario. We deal with the corresponding boundary conditions of a model detector such as a mirror, and present a consistent solution to the spectral problem in the context of Sturm-Liouville theory. The solution involves only one type of Hankel function with imaginary argument and index \footnote{The $-1/x^2$ potential happens to possess localized $L^2(\mathbf{R_+})$ solutions for continuos energies. Their quantization needs an additional condition.}.

The familiar Galilean case corresponds to an accelerated wall against a stationary particle, thus explaining the quantization of the energy spectra by a constant and fictitious gravitational field that produces bouncing modes \cite{colella, greenberger, ryder}. In the relativistic case, the existence of continuous or discrete wave numbers --surprisingly, both cases localized by $-1/x^2$ potential wells in one dimension-- comes from the Christoffel symbols appearing in the field equations. Their emergence takes place when transitioning from the Minkowski space to the famous Rindler metric. Such terms are known as quantum \textit{anomalous} interactions with a fall-to-the-origin behavior \cite{Falltotheorigin}. Their strong singularity forbids the application of some familiar theorems (Sturm-Liouville, Kato-Rellich \cite{kato}) valid for regular, self-adjoint wave operators. This interesting potential is notable in quantum mechanics for providing an exactly solvable Helmholtz problem and for leading to a homogeneous Hamiltonian operator of degree $-2$, which preserves its form under scale transformations \cite{Moroz, SadurníCastillo}. Unlike the Coulomb potential $-1/x$ , the fictitious interaction in question lacks a ground state, preventing spectrum quantization if the vertex of Rindler's wedge is included in the domain. To adequately treat this potential, an additional boundary condition must be introduced, but this arises naturally when a material wall is pushing the particles or fields in an accelerated manner, i.e., a piston.

A natural force that behaves in this anomalous manner would be an idealized electric dipole, but there are no point-like electromagnetic or gravitational sources that lead to such a strong singularity. Instead, the effect in question is described by a coordinate transformation acting on the observer's position and on the material wall, modeled by a movable boundary. This kind of potential has a geometric interpretation explained by the FitzGerald contraction applied to relativistic observers with uniform acceleration. For example, a space-like rod with two ends, front and rear, when uniformly accelerated, must locally adjust its length such that the rear end catches up with the front in order to maintain a constant acceleration. Thus, for consistency with relativistic kinematics, the deformation of the rear end must be greater than the front and this, in turn, produces a singular point $x=0$ where the rod's length vanishes.

The equivalent non-relativistic problem is interpreted as a fall-to-the origin effect described by the width $\langle r^2 \rangle$ of a wavepacket initially centered at the origin. Using the virial theorem, it is found that for a homogeneous potential $V = V_0 r^n$, with $n = -2$, 
\begin{equation}
\frac{d^2}{dt^2} \langle r^2 \rangle = -\langle H \rangle, 
\end{equation} with $\langle H \rangle$ being a constant. Therefore, the acceleration of $\langle r^2 \rangle$ is negative and constant. Thus it is demonstrated that the wave packet collapses in a finite time using this equation of motion.

In this work, we first explore the phenomenon in the non-relativistic case, using the Schrödinger equation subjected to the uniformly accelerated Dirichlet boundary condition (Section \ref{Non-relativistic case}). The complete family of relativistic coordinate transformations that represent uniformly accelerated observers is explained in Section \ref{Family of Rindler Coordinates}. We then proceed with the Klein-Gordon equation for a scalar field (Section \ref{Klein-Gordon equation}) and the Maxwell vector field (Section \ref{Equation for the classical Maxwell field}). For photons inside a piston, we include mirrors and polarizers in Section \ref{Photons against a wall and inside a piston}. The remaining pair of components, longitudinal and scalar, are addressed in \ref{General Equations for the Field Tensor}. The mathematical properties of spectral anomalies are indicated in Section \ref{Mathematical properties of the anomalous wave operator}. Our efforts pay dividends in the evaluation of transition overlaps by menas of the Bogoliubov transformation from inertial to non-inertial field operators; this is done in Section \ref{sec:physical}. We conclude in Section \ref{Conclusions}.

\section{Non-relativistic case}
\label{Non-relativistic case}
First, we study the non-relativistic case. Unlike the relativistic case, there is no singularity at $x=0$, and the wave function is regular for all values of $x$. Therefore, the theory of self-adjoint operators (Sturm-Liouville) is always valid. In this system, there is no quantization of the spectrum without the existence of the Dirichlet boundary condition. The resulting eigenvalues will be bounded only from below, as the problem shall be equivalent to a triangular potential well.

We begin our analysis with the Schrödinger equation for a free particle (with $\hbar=1, \; m=1/2$), subjected to a moving boundary condition, i.e. a uniformly accelerated Dirichlet wall:
\begin{equation}
\begin{aligned}
\psi\left(at^2/2,t\right)=0.
\end{aligned}
\end{equation}
We apply a change of variables for an accelerated frame along classical trajectories:$
t^{\prime}=t, \; x^{\prime}=x-at^2/2.$ The transformed equation will be
\begin{equation}
\begin{aligned}
\left[-\frac{\partial^2}{\partial x^{\prime 2}} +iat^{\prime}\frac{\partial}{\partial x^{\prime }}-i\frac{\partial}{\partial t^{\prime }}\right] \phi(x^{\prime},t^{\prime})=0,
\label{ecuación Sc}
\end{aligned}
\end{equation}
where $\phi(x^{\prime},t^{\prime})\equiv\psi(x^{\prime}+at^{\prime 2}/2,t^{\prime})$, thus achieving a static boundary condition $\phi(x^{\prime}=0,t^{\prime} )=0, \;\forall \, t^{\prime}$. This transformation entails a new wave function $\phi$ and the corresponding Jacobian is unity, therefore the normalization in $x'$ is preserved.
In this reference frame the boundary condition is static, so the particle moves toward the wall. The expression (\ref{ecuación Sc}) is conveniently rewritten as
\begin{equation}
\begin{aligned}
\left[-\left( \frac{\partial}{\partial x^{\prime}} -i \frac{at^{\prime}}{2}\right)^2 - \frac{a^2 t^{\prime 2}}{4} -i\frac{\partial}{\partial t^{\prime}}\right] \phi(x^{\prime},t^{\prime})=0,
\end{aligned}
\end{equation}
which allows for a gauge transformation
\begin{equation}
\begin{aligned}
 \frac{\partial}{\partial x^{\prime}} - i \frac{at^{\prime}}{2}= e^{i at^{\prime}x^{\prime}/2}  \frac{\partial}{\partial x^{\prime}} e^{-i at^{\prime}x^{\prime}/2}.
\end{aligned}
\end{equation}
We arrive at a new Schr\"odinger operator associated to the equation
\begin{equation}
\begin{aligned}
\left[- \frac{\partial^2}{\partial x^{\prime 2}}- \frac{a^2 t^{\prime 2}}{4}-i\frac{ \partial}{\partial t^{\prime}}+ \frac{a x^{\prime}}{2} \right] \tilde{\phi}(x^{\prime},t^{\prime})=0,
\label{ecuación separable}
\end{aligned}
\end{equation}

where $\tilde{\phi}(x^{\prime},t^{\prime})\equiv e^{-iat^{\prime}x^{\prime}/2}\phi(x^{\prime},t^{\prime})$. This is a separable equation that can be solved with the following \textit{ansatz}: $\tilde{\phi}(x^{\prime},t^{\prime})=U(x^{\prime})V_0 e^{-ia^2 t^{\prime 3}/2-i\varepsilon t^{\prime}}$ and the boundary condition $U(0)=0$. We rescale our variables $u = x^{\prime} \left( a/2 \right)^{1/3}$, $\tilde{U} = 2U/a$ and $\tilde{\varepsilon} = \epsilon (a/2)^{-2/3}$, such that the spatial part satisfies a dimensionless Airy equation:
\begin{equation}
\begin{aligned}
\left( -\frac{d^2}{d  u^2}+u-\tilde{\varepsilon} \right) \tilde{U}=0.
\end{aligned}
\end{equation} 
Obviously, the solution to this equation is the Airy function $\text{Ai}(z)$. The condition at $x\rightarrow +\infty$ rules out functions of the type $\text{Bi}(z)$ due to their divergent integrals. Therefore, the general solution in $\mathbf{R_+}$ is
\begin{equation}
\begin{aligned}
\tilde{U}(u)= U_0 \text{Ai}(u-\tilde{\varepsilon}).
\end{aligned}
\end{equation}
Finally, the boundary condition $\tilde{U}(0)=0$ leads to the familiar quantization:
\begin{equation}
\begin{aligned}
\text{Ai}(-\tilde{\varepsilon})=0 \quad \rightarrow\quad \varepsilon_n=\alpha_n, 
\end{aligned}
\end{equation}
\begin{equation}
\begin{aligned}
\varepsilon_n = \left(\frac{a}{2}\right)^{2/3} \alpha_n,
\end{aligned}
\end{equation}
where $\alpha_n$ are the well-known roots of the Airy function. Furthermore, using the \textup{WKB} method, we find an approximation for the energy levels based on the Sommerfeld condition, which in this case takes the form:
\begin{equation}
\varepsilon_n \propto\left(n+\frac{1}{4}\right)^{2 / 3}.
\end{equation}
Contrary to the problem of a particle in a constant force field, our problem is divided into two regions separated by the presence of the Dirichlet wall: First, we have a region $x<0$, where the energy is \textit{continuous} and the solutions are not square-integrable. Second, the region $x>0$ in the direction of accelerated motion, allows bound states and the solutions are $L_2(\mathbf{R}_+)$.

The latter states are neither equally spaced nor accumulated, but their distance decreases according to $n^{2/3}-(n-1)^{2/3}$. We can interpret that the acceleration of the wall “compresses” the wave in the direction of motion, i.e. its action is similar to a piston. Consequently, the uniformly accelerated boundary condition induces the quantization of the energy spectrum on the right-hand side of the wall. These mathamtical feautures shall be important in the rest of the paper, but as it is now obvious, a relativistic treament that avoids superluminal velocities is indispensable.

\section{Family of Rindler Coordinates}
\label{Family of Rindler Coordinates}
\begin{figure}[h]
\centering
\includegraphics[scale=.25]{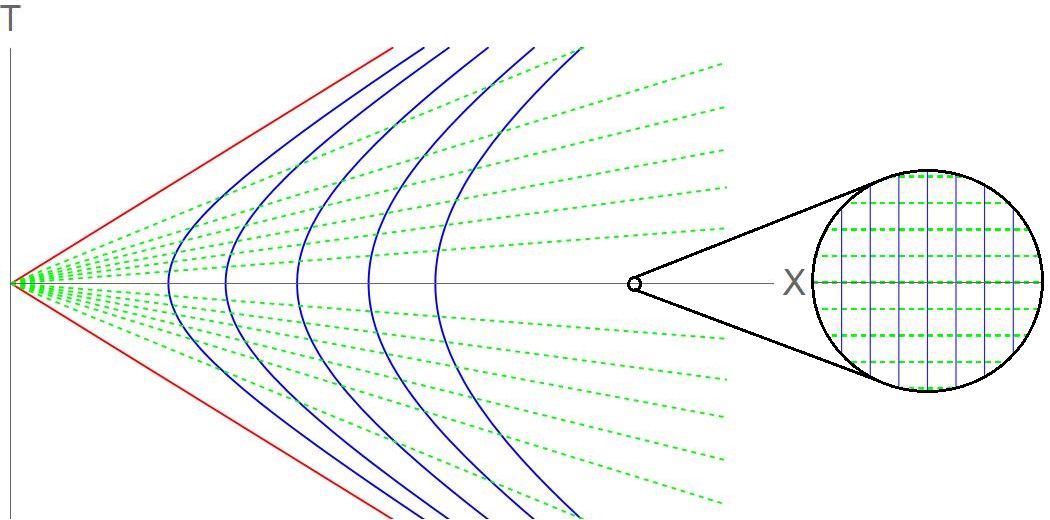}
\caption{\label{rindler}Spacetime diagram for a Rindler observer. The red lines represent the Rindler wedge and the blue hyperbolas correspond to uniformly accelerated observers.}
\end{figure}

To find trajectories with uniform acceleration that are compatible with relativity, the dynamics must comply with a local Lorentz transformation between the velocities in the inertial frame $v_\eta$ and the accelerated frame $v_\beta$. Such a relation is given by
\begin{equation}
v_\eta (\tau)= \Lambda^{\beta}_{\eta}(\tau)v_\beta.
\end{equation}
The Lorentz transformation $\Lambda^{\beta}_{\eta}$ may depend explicitly on time $\tau$, indicating that, from the point of view of an inertial observer, an accelerated object is described by a nonconstant $v_{\eta}(\tau)$. Therefore, the tensor $\Lambda$ produces an evolution of the 4-vector $v_{\beta}(t_0)$, leading to the following equation of motion
\begin{equation}
\alpha_\eta \equiv \frac{d v_\eta}{ d \tau} = \frac{d\Lambda_\eta ^{\beta}}{ d \tau} v_\beta(t_0).
\end{equation}
Taking the hyperbolic representation of the Lorentz transformation 
\begin{equation}
\Lambda =
\begin{pmatrix}
\cosh\phi & -\sinh\phi \\
-\sinh\phi & \cosh\phi
\end{pmatrix},
\end{equation}
and computing the derivative, the equation of motion takes the form
\begin{equation}
\binom{d v_0/d \tau}{d v_1/d\tau}=-\frac{d \phi}{d \tau}\binom{v_1}{v_0}.
\label{sistema de ecuaciones}
\end{equation}
Considering that the acceleration $\alpha_{\eta}$ must have uniform norm, i.e. $\alpha_\nu \alpha^\nu = -\alpha^2 = \text{constant}$, in analogy with the nonrelativistic case, 
the solutions of (\ref{sistema de ecuaciones}) are 
\begin{equation}
\begin{gathered}
v_0=A \cosh \phi+B \sinh \phi, \\
v_1=-A \cosh \phi-B \sinh \phi,
\end{gathered}
\end{equation}
as long as the angular frequency is chosen as constant, e.g. $d\phi / d\tau =\bar \omega$, so $\phi = \bar{\omega} \tau$ and $A^2 - B^2 = \alpha^2 / \bar{\omega}^2$. The solutions are expressed as
\begin{equation}
\begin{gathered}
v_0=A \cosh \phi \pm \sqrt{A^2-\frac{\alpha^2}{\bar{\omega}}} \sinh \phi, \\
v_1=-A \cosh \phi\mp \sqrt{A^2-\frac{\alpha^2}{\bar{\omega}}} \sinh \phi.
\label{soluciones para v}
\end{gathered}
\end{equation}
Finally, by integrating (\ref{soluciones para v}), we obtain the coordinates of a uniformly accelerated object from an inertial frame 
\begin{equation}
\begin{aligned}
& \binom{x_0(\tau)}{x_1(\tau)}=x(0)+\frac{1}{\bar{\omega}}\binom{\mp \sqrt{A^2-\alpha^2/\omega}}{A} -\frac{1}{\bar{\omega}}\left(\begin{array}{cc}
\cosh \phi & -\sinh \phi \\
-\sinh \phi & \cosh \phi
\end{array}\right)\binom{\mp \sqrt{A^2-\alpha^2/\omega}}{A},
\end{aligned}
\end{equation}
This can be regarded as the biparametric family of Rindler transformations. If we choose $\bar{\omega} = -\alpha/c$, $x(0)=(0,0)^{\rm T}$, and consider that
\begin{equation}
 \text { }\binom{t}{x^1}=\binom{t^{\prime}}{\left(x^1\right)^{\prime}},
\end{equation}
the coordinate transformation is simplified to a more familiar form that only includes the application of the $\phi$-dependent matrix to the old coordinate vector. In the limit $\alpha \rightarrow 0$, we recover the constant Lorentz transformation between vectors. With this in mind, a possible definition of transformed coordinates, given by Rindler, is obtained by applying the reversed Lorentz transformation above to a vector the contains only the proper length $x$, i.e.
\begin{equation}
T=x \sinh \alpha t, \quad X=x \cosh \alpha t ,
\end{equation}
which gives rise to the following components of the metric
\begin{equation}
g_{00}=\alpha^2 x^2, \quad g_{i j}=-\delta_{i j}, \quad g_{0 i}=0.
\end{equation}
It is important to note the singularity that emerges at $x=0$ (wedge) in contrast with the nonrelativistic case; this shall affect the regularity of wave operators. Finally, the limit when the acceleration approaches zero is reached only if $x$ is taken in the region $x \rightarrow 1/\alpha$, which corresponds to the asymptotically flat portion of Rindler space, as shown in Fig. \ref{rindler}.

\section{Klein-Gordon equation}
\label{Klein-Gordon equation}
The extension of the free Klein-Gordon equation in flat space to curved space requires the replacement of the ordinary derivative operator $\partial_\mu$ by its covariant form $\nabla_\mu$, with the recipe
\begin{equation}
\left(\nabla_\mu \nabla^\mu+m^2\right)\phi=\left(\partial_\mu \partial^\mu+\Gamma_{\mu \lambda}^\mu \partial^\lambda+m^2\right)\phi=0.
\end{equation}
The scalar curvature $R$ is usually included as an extra term, but in our specific problem $R\equiv0$. If we now use the property of the trace of the Christoffel symbol, i.e., $\Gamma_{\mu \lambda}^\mu = \partial_\lambda(\sqrt{|g|}) / \sqrt{|g|}$, we arrive at
\begin{equation}
0=  {\left[\left(\partial_\lambda+\frac{1}{2} \partial_\lambda(\log \sqrt{|g|} \mid)\right) \right.} \left.\left(\partial^\lambda-\frac{1}{2} \partial^\lambda(\log \sqrt{|g|})\right)+m^2\right]|g|^{1 / 4} \phi.
\end{equation}
For the case of Rindler space, $|g|=\alpha^2 x^2$, and we find $\left[\square+M^2(x)\right] \psi=0,$ where $M^2(x) = m^2 - 1/(2x)^2$ is interpreted as an effective mass, and the rescaled wave fucntion is $\psi = |g|^{1 / 4} \phi$, which satisfies proper normalization conditions in curved space. Such normalization comes from the well-known Klein-Gordon inner product, i.e. $\left\langle f, h \right \rangle = i \int dx^{3} \sqrt{|g|}(f^{*}\partial_{t}h-h\partial_{t} f^{*})$, which reduces to the $L_2$ norm for stationary states of positive energy. Using the definition $\square \equiv \partial_\mu g^{\mu \nu} \partial_\nu$, the KG equation can be written as
\begin{equation} 
\left[g^{00} \partial_0^2 - \nabla^2 + M^2(x)\right] \psi = 0. 
\label{ecuación KG con g00} \end{equation}
 Now we show that the solution is separable as $ \psi=e^{ \pm i \omega t} f(x) e^{i \mathbf{k}_{\perp} \cdot \mathbf{x}_{\perp}},$ where we define a transverse wave vector $\mathbf{k}_{\perp}=(k_y, k_z)$ and $\mathbf{x}_{\perp} = (y, z)$. Substituting into (\ref{ecuación KG con g00}), we get
\begin{equation}
\left[\frac{\omega^2}{\alpha^2 x^2}+\frac{d^2}{d x^2}+\frac{1}{4 x^2}-\left(k_{\perp}^2+m^2\right)\right] f(x)=0.
\label{bessel KG}
\end{equation}
This is equivalent to a Bessel equation expressed in terms of a symmetric differential operator, but when viewed as an eigenvalue problem, it corresponds to a one-dimensional stationary Schr\"odinger equation with anomalous potential $-1/x^2$ and energy $-(k_{\perp}^2+m^2)<0$. It should be noted that the previous non-relativistic case can be recovered in a special limit. To do this, we need $\alpha \rightarrow 0$ and $x\rightarrow 1/ \alpha+u$, where $u$ is small. In other words, one must avoid the Rindler singularity. It should be noted, though, that in both cases the eigenstates to the right of the moving wall are $L^2(\mathbf{R}_+)$, but the spectra are qualitatively different; here the quantized spectrum is \textit{negative}, the potential well is unbounded below and requires an additional boundary condition.

In general, the solution to eq. (\ref{bessel KG}) is:
\begin{equation}
f(x) =A \sqrt{x} H_{2 i \omega / \alpha}^{(1)}\left(i x \sqrt{k_{\perp}^2+m^2}\right) +B \sqrt{x} H_{2 i \omega / \alpha}^{(2)}\left(i x \sqrt{k_{\perp}^2+m^2}\right),
\end{equation}
where the factor $\sqrt{x}$ is the Jacobian $|g|^{1/4}$ coming from the metric, which is necessary to properly define the inner product, and it is one of the requirements to produce a self-adjoint operator of the form
\begin{equation}
   \mathcal{H} \equiv -\frac{d^2}{dx^2} - \frac{(\omega/\alpha)^2 + 1/4}{x^2}. 
   \label{selfad}
\end{equation}
If we take into account the boundary condition $|f|\rightarrow 0$ as $x \rightarrow \infty$, we discard the term $B$ as it grows exponentially. Then, the domain of the operator (\ref{selfad}) is $L_2(\mathbf{R}_+)$ if $A$ is kept, and this can be verified from the definition of the Hankel function for imaginary argument and index \cite{ComputationHankel} using the Schl\"afli integral.   

In some reviews of the Unruh effect and mathematical treatments of the quantum potential $-1/x^2$, one can hardly find physical considerations in the discrimination of adequate solutions. For example, \cite{Crispino} provides a different function in terms of the modified Bessel function $K_{\nu}$ and an exponential argument, while \cite{Berry} also provides the solution in terms of $K_\nu$, but keeps the full real line as a valid domain for the integrals in the inner product. This is proposed without discarding the second Hankel function in its definition; as already stated, the latter grows exponentially for imaginary argument. We note here that this problem can be stated as a relation between $K_{\nu}$ and $H_{\nu}^{(1)}$, which we explain in eq.(\ref{modifiedbessel}), in section VII. In fact, when these functions are confounded, computations of transition amplitudes in the Unruh effect could lead to unfortunate divergent integrals.

Furthermore, the energies are not yet discretized. If we impose the boundary condition of the wall at $x=x_m>0$, the energies will be quantized in the following form:
\begin{equation}
\begin{aligned}
& H_{2 i \omega / \alpha}^{(1)}\left(ix \sqrt{k_{\perp}^2+m^2}\right)=0, \\
& x_m \sqrt{k_{\perp}^2+m^2}=r_n(\nu), \quad n=0,1,2, \ldots
\end{aligned}
\end{equation}
where $r_n(\nu)$ are the (real) roots of the Hankel function with imaginary argument and imaginary index.

Thanks to the boundary condition, we avoid the singularity of the wave operator; therefore, the necessary conditions for the application of the Sturm-Liouville theorem for self-adjoint operators are satisfied. It should be clarified that, unlike the Galilean case, the boundary condition cannot be placed at $x=0$, which forces the observer to avoid the Rindler wedge. In the following section, we move on to the case of photons. We will analyze the mathematical properties of these solutions in more detail in Section VII.

\section{Equation for the classical Maxwell field}
\label{Equation for the classical Maxwell field}
In curved spacetime, the Maxwell equations are
\begin{equation}
\nabla^\mu F_{\mu \nu}=4 \pi j_\nu, \quad \nabla_{[\mu} F_{\nu \rho]}=0.
\label{Maxwell equations}
\end{equation}
The electromagnetic field tensor in terms of the vector potential is $ F_{\mu \nu} = \nabla_\mu A_\nu - \nabla_\nu A_\mu = \partial_\mu A_\nu - \partial_\nu A_\mu$. 
Substituting into the first equation of (\ref{Maxwell equations}) for the case $j_\nu =0$, and using the Lorenz gauge $\nabla_\mu A^{\mu} = 0$, obtains the modified wave equation
\begin{equation}
\nabla^\nu \nabla_\nu A_\mu-R^\rho{ }_\mu A_\rho=0,
\label{Maxwell equation in curved space}
\end{equation}
where $R^{\rho}_{\mu}$ is the Ricci tensor. In principle, this term couples to \textit{polarization} and has the capability of rotating the physical fields $\mathbf{E}, \mathbf{B}$. For the Rindler metric, this can be greatly simplified; the Christoffel symbols that do not vanish identically are $\Gamma^1_{00} = \alpha^2 x$, $\Gamma^0_{10} = \Gamma^0_{01} = 1/x$. With this, it can be verified that $R^\rho{ }_\mu = 0$. By developing (\ref{Maxwell equation in curved space}), we arrive at
\begin{equation}
g^{\alpha \nu} \nabla_\alpha \nabla_\nu A_\mu=\partial^\nu\left(\partial_\nu A_\mu-\Gamma_{\nu \mu}^\sigma A_\sigma\right)-g^{\alpha \nu} \Gamma_{\alpha \nu}^\lambda\left(\partial_\lambda A_\mu-\Gamma_{\lambda \mu}^\sigma A_\sigma\right)-g^{\alpha \nu} \Gamma_{\alpha \mu}^\lambda\left(\partial_\nu A_\lambda-\Gamma_{\nu \lambda}^\rho A_\rho\right) .
\label{reducedmaxwell}
\end{equation}
It is observed that we will have different differential equations for the components of the vector potential ($\mu=0,1,2,3$) due to their distinct coupling with the Christoffel symbols.

In the following section, we will focus on the components perpendicular to the motion, namely $A_2$ and $A_3$. To achieve this, a boundary condition is introduced—a mirror—which filters out certain waves while reflecting those contributing to the Unruh effect. Our physical setup will consist of a polarizer that decouples the $A_2$ and $A_3$ components, thereby simplifying the differential equations.

\section{Photons against a wall and inside a piston}
\label{Photons against a wall and inside a piston}

\begin{figure}[h] 
    \centering 
    \includegraphics[width=0.6\textwidth]{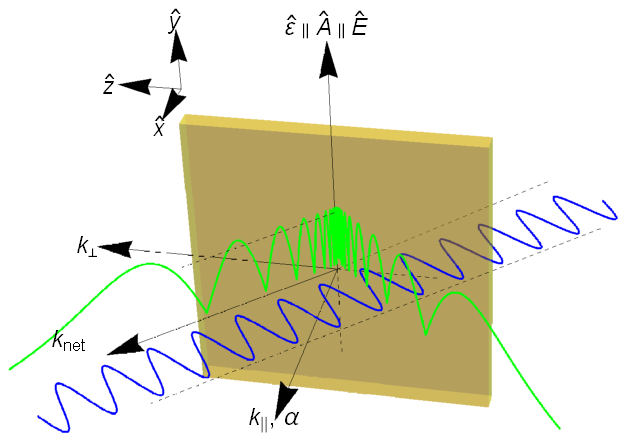}
    \caption{Accelerated polarizer in the direction of the $\hat{x}$ axis with perfect reflectivity along $\hat{x}$.}
    \label{Polatizador}
\end{figure}

In what follows, we adopt a coordinate system defined by an infinite plane perpendicular to the $x$ axis, that models a very large mirror reflecting the $y$ component of the field (linear polarization). This is shown in Fig.\ref{Polatizador} in the form of a Dirichlet wall moving with constant acceleration. A finer set of solutions can be provided if the mirror is finite, as diffraction would contribute with Fresnel functions, but for the moment let us assume that all wavelengths are smaller than the obstacle's dimensions.

We use the notation $\mathbf{k}_{\text{net}} = \mathbf{k}_{\perp} + \mathbf{k}_{\|}$ for the total wave vector, $\mathbf{k}_{\perp} \cdot \mathbf{x}$ is in the direction perpendicular to the motion of the mirror, while $\mathbf{k}_{\|} = \hat{x} k_{\|}$ is in the parallel direction. Then, a uniformly accelerated beam splitter, as in Fig. \ref{Polatizador}, with perfect reflectivity, ensures that only the components $A_2$ and $A_3$ of the field can interact with it. In this way, one can ignore photons with other polarizations, and in our equation $A_1 = 0$ is a sensible choice. On the other hand, if we also work in the radiation gauge, we can take $A_0 = 0$. This will significantly simplify the equations to be solved.

With the use of the polarizer, the field is reduced to
\begin{equation}
\begin{gathered}
A_2=e^{-i \omega t} e^{i \mathbf{k}_{\perp} \cdot \mathbf{x}} F_2(x), \\
A_3=e^{-i \omega t} e^{i \mathbf{k}_{\perp} \cdot \mathbf{x}} F_3(x), \\
\mathbf{k}_{\perp} \cdot \mathbf{A}=0.
\end{gathered}
\label{componentes del campo}
\end{equation}

Substituting the separable solution ($\mu = 2, 3$) the Maxwell wave equation (\ref{Maxwell equation in curved space}) finally boils down to the massless Klein-Gordon equation in curved space (\ref{bessel KG}) for each component of $A$, that is,
\begin{equation}
\left(\Box - \frac{1}{4x^2} |g|^{1/4} \right)A_\mu = 0.
\label{para a}
\end{equation}
In this way, the operator in (\ref{selfad}) appears once more and we conclude that the function $\psi = |g|^{1/4} F(x)$ must be a solution to the equation
\begin{equation}
\left[-\frac{d^2}{d x^2}-\frac{(\omega / \alpha)^2+1 / 4}{x^2}\right] \psi=-k_{\perp}^2 \psi .
\label{operador autoadjunto}
\end{equation}

\begin{figure}[h] 
    \centering 
    \includegraphics[width=0.7\textwidth]{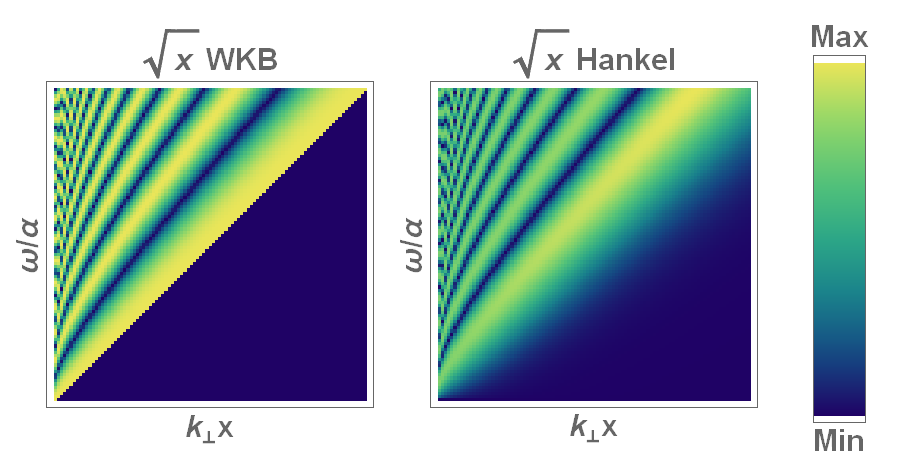}
    \caption{Probability density for the  wave function, comparison between the exact result and the WKB approximation.
} 
    \label{solución completa}
\end{figure}

 As before, the solution involves Hankel functions of type $1$ and $2$, but once again, the functions $H^{(2)}_\nu$ are discarded due to their exponential growth as $x \to \infty$. Therefore, $H^{(1)}_\nu(z)$ is kept, where $\nu \equiv i\omega/\alpha$ and $z \equiv ik_\perp x$, that is, with imaginary argument and index. We emphasize that $\sqrt{x}H^{(1)}_\nu(ik_\perp x)$ is the solution for the eigenvalue problem with a self-adjoint operator, as the Jacobian regularizes the wave function at $x=0$, as shown on the right side of the Fig.\ref{solución completa}.

It is interesting that in the case of $k_\perp = 0$, the solutions obtained correspond to scattering states that do not decay as $x \to \infty$, meaning they are not square-integrable solutions. Therefore, the angle of incidence also influences the quantization of the spectrum. But when considering all the radiation that interacts with the mirror, these do not contribute to the calculation.

\begin{figure}[h] 
    \centering 
    \includegraphics[width=0.7\textwidth]{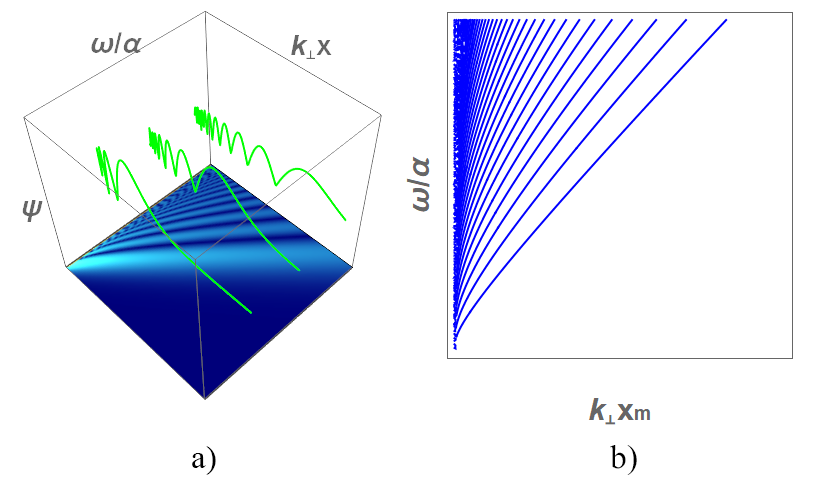}
    \caption{a) Hankel functions for various values of $\omega/\alpha$ (green curves) and density plot for $|H^{(1)}_{iy}(ix)|^2$. b) Frequency quantization for the polarizer as a boundary condition.} 
    \label{Hankel}
    \end{figure}

If we now take into account the Dirichlet boundary condition at $x = x_m > 0$, which results from the presence of the accelerated mirror, we arrive at the quantization condition for the solution:
\begin{equation}
H_{i \omega / \alpha}^{(1)}\left(i k_{\perp} x_{\mathrm{m}}\right)=0,\quad  k_{\perp} x_{\mathrm{m}}=r_n(\nu), \quad n=0,1,2, \ldots ,
\label{cuantización de H}
\end{equation}
where $r_n(\nu)$ are the real roots of $H^{(1)}_\nu(\zeta)$. In Fig.\ref{Hankel} a) we show a graph of these functions for various values of the argument and the index, while in b) we have the quantization of the frequencies when we introduce the boundary condition of the polarizer.

In Fig.\ref{solución completa}, we present a comparison: on the right, the exact solution of the Hankel functions, and on the left, a plot for the $\mathrm{WKB}$ approximation, in which the integral representation for the Hankel functions has been used (\cite{Manual}, p. 915, sec. (8.421), formula 8):
\begin{equation}
\begin{aligned}
H_\nu^{(1)}(x z) & =\frac{z^\nu e^{-i \nu \pi / 2}}{i \pi} \int_0^{\infty} d t e^{i x\left(t+z^2 / t\right) / 2} t^{-\nu-1} \\
& =\frac{1}{i \pi} \int_{-\infty}^{\infty} d u \quad e^{-\nu u+i|z| x \sinh u} \\
& \approx \sqrt{\frac{8 i \pi}{\sqrt{|\nu|^2-(|z| x)^2}}} \cos \left[|\nu| \operatorname{arccosh}\left(\frac{|\nu|}{|z| x}\right)\right. \\
& \left.-\sqrt{|\nu|^2-(|z| x)^2}-\pi / 4\right] \Theta\left(\left|\frac{\nu}{z}\right|-x\right) .
\label{WKB Aproximación}
\end{aligned}
\end{equation}
The stationary phase approximation was used to obtain the result. It is a good approximation if $\alpha << \omega$ and $1/x_m << |\vec{k}_\perp|$, and for the approximation to hold, $|\vec{k}_\perp| < \omega$. A large $|\vec{k}_\perp|$ implies a small wavelength $\lambda$, that is, a semiclassical limit. Therefore, $\mathrm{WKB}$ holds if $x > 1/\alpha$, avoiding that the variation of the potential $|V^{\prime}(x)|$ is too large.

\begin{figure}[h] 
    \centering 
    \includegraphics[width=0.6\textwidth]{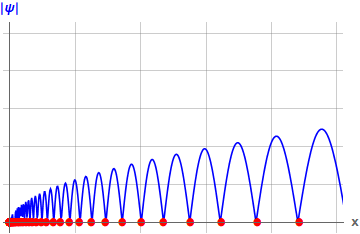}
    \caption{A comparison of the zeros obtained with the approximation (\ref{aproximación de ceros}), red dots, and the exact graph of the wave function $\psi(x)=\sqrt{x}H^{(1)}_\nu(ik_\perp x)$, blue curves, is shown.} 
    \label{grafica comparación de ceros}
    \end{figure}

With the result (\ref{WKB Aproximación}), we can obtain an approximation of the energy levels (zeros of the Hankel function). If we take $\nu \gg |z|x$ and knowing that $\text{arc}\cosh(u) \approx \ln(2u)$ when $u \gg 1$, we obtain the equation: 
\begin{equation}
x_n=\frac{|\nu|}{2|z|} e^{-\frac{1}{|\nu|}\left(n+\frac{3}{4}\right) \pi}.
\label{aproximación de ceros}
\end{equation}
In Fig.\ref{grafica comparación de ceros}, the comparison of the result with the roots of the exact wave function is shown. This semiclassical approximation agrees with the exponential ladder in \cite{SadurníCastillo}, where the spectrum is derived for a fractal potential that scales as $-1/x^2$.

\subsection{The piston}

\begin{figure}[h] 
    \centering 
    \includegraphics[width=0.6\textwidth]{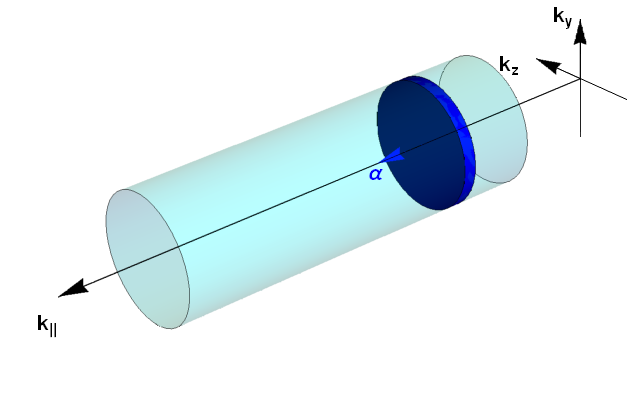}
    \caption{Geometry of a piston: a waveguide containing a plunge made of reflective materials and externally covered by a conductor. The plunge (blue) is accelerated.
} 
    \label{guia de onda 1}
\end{figure}

While our previous system required an infinite mirror, it is also possible to use a finite wall with a planar shape --e.g. a rectangle or a circle. A \textit{piston} is defined by a cylindircal cavity whose axis coincides with $x$, as seen in Fig. \ref{guia de onda 1}, and the field is compressed by the acceleration of such a mirror or \textit{plunger}. The boundary conditions must comply with the fact that $\mathbf{A} \, || \, \mathbf{E}\,||\,\hat y$ and satisfies Dirichlet or Neumann conditions according to whether the cylinder walls are parallel or perpendicular. Twisted waveguides may constitute an attractive possibility, due to their binding capabilities, \cite{dacosta, dacosta2, Goldstone}, but for illustrative purposes we can deal with a rectangular shape such that $A_1 = F(x)\sin(l \pi z/L_z)\cos(m \pi y/L_y)$. This quantizes $k_{\perp}$ from the outset, and the equation for $F(x)$ is the same as (\ref{para a}) and (\ref{operador autoadjunto}), except that $k_{\perp}=\pi \sqrt{(l/L_z)^2 + (m/L_y)^2}$ is fixed. Therefore, the same solutions to (\ref{operador autoadjunto}) can be applied, but at the end of the process one must solve for $\omega$ in terms of $n, l, m$. Semiclassically, instead of the more complicated (\ref{cuantización de H}), we employ the implicit relation

\begin{align}
2 \pi x_{\rm{m}} \sqrt{ (l/L_z)^2 + (m/L_y)^2 }  = \frac{\omega_{n,l,m}}{\alpha}e^{-\omega_{n,l,m} \pi(n + 3/4)/\alpha}   
\end{align}
 that leads to the approximation ($n$ large)

\begin{align}
\omega_{n,l,m} \approx \frac{\alpha}{\pi(n+3/4)} \log\left[\frac{1}{2 \pi x_{\rm{m}} \sqrt{ (l/L_z)^2 + (m/L_y)^2 }}\right].    
\end{align}

If instead, we propose an arbitrary transverse shape (for example a stadium billiard) the solutions for the transverse functions in $y,z$ could be irregular or even chaotic \cite{Stockmann,Bunimovich,Legrand}. In the present work it is not indispensable to deal with these more complicated geometries, but future statistical analysis may be of interest.

\subsection{Coalescence of energy levels}
\begin{figure}[h] 
    \centering 
    \includegraphics[width=0.6\textwidth]{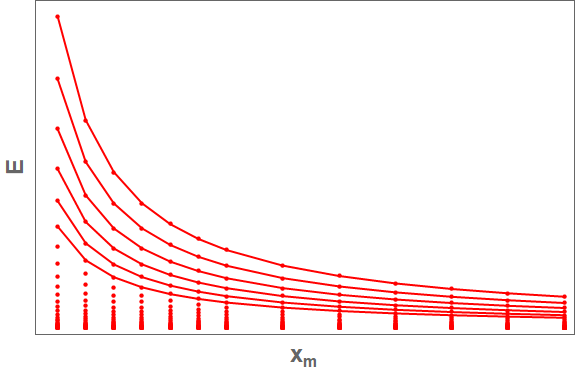}
    \caption{Energy levels as a function of the boundary condition position $x_m$, given by (\ref{Energía en función de xm}), it is observed how the discrete energy levels tend to a continuum.
} 
    \label{Draft1.1}
\end{figure}

Energy or frequency quantization disappears in the limit of low acceleration. This is seen as a coalescence of energy levels as the boundary condition approaches the origin, which is observed in Fig.\ref{Hankel} b). This can be deduced using the integral representation in (\ref{WKB Aproximación}), analyzing the argument of the cosine.  Replacing $x \mapsto ix$, $\nu \mapsto i\nu$, and using the change of variable $\tau = x t$, leads to
\begin{equation}
\begin{aligned}
&x^{i\nu} I\equiv x^{i \nu} \int_0^{\infty} d\tau \, \tau^{-i \nu-1} e^{-\tau-(x z)^2 / \tau}.
\end{aligned}
\end{equation}
The integral $I$ is a function of the product $xz$. The real part of the function is
\begin{equation}
\begin{aligned}
\sqrt{x} \operatorname{Re}\{I\} & =\sqrt{x}[\cos (\nu \log xz) \operatorname{Re}\{I\}+\sin (\nu \log xz) \operatorname{Im}\{I\}], \\
& =\sqrt{x} \sqrt{\operatorname{Re}^2 \{I\}+\operatorname{Im}^2\{I\}} \cos \left(\nu \log xz+\gamma_v\right),
\end{aligned}
\end{equation}
where $\gamma_\nu =-\arctan \left[\operatorname{Im}\{I\}/ \operatorname{Re}\{I\} \right]$ is a phase. Now, setting $\operatorname{Re}\{I\}=0$ to find the roots $r_n =x_m z_n$, we arrive at the expression 
\begin{equation}
\begin{aligned}
& \nu\log \left(x_m z \right)+\gamma_v=\pi(n+1 / 2), \quad n \in \mathbb{Z},  \\
\end{aligned}
\end{equation}
which leads to
\begin{equation}
z_{n+1}-z_n \cong \frac{\partial z}{\partial n}=\left(\frac{\pi}{x_m \nu} \right) e^{\left[\pi(n+1 / 2)-\gamma_\nu\right]/\nu}.
\label{Energía en función de xm}
\end{equation}
From this expression, it is verified that if $x_m \rightarrow 0$ and $n \rightarrow -\infty$, the sequence of levels tends to a continuum. This is represented in Fig.\ref{Draft1.1}.

\section{General Equations for the Field Tensor $F^{\mu \nu}$}
\label{General Equations for the Field Tensor}
In this section, we focus on finding the solution for the remaining components of the vector potential: the scalar part $A^0$ and the longitudinal component $A^1$. For this purpose, we employ the equation for the electromagnetic tensor. From (\ref{Maxwell equations}), the equation for a Rindler metric is:
\begin{equation}
\partial_{\mu}\left(x F^{\mu \nu}\right)=0 .
\label{ecuación de Maxwell para F}
\end{equation}
Expanding for $\nu=i=2,3$ yields:
\begin{equation}
\frac{1}{\alpha^2 x^2} \frac{\partial F_{02}}{\partial t} - \frac{\partial F_{12}}{\partial x} - \frac{\partial F_{32}}{\partial z} - \frac{1}{x} F_{12} = 0, \quad \;
\frac{1}{\alpha^2 x^2} \frac{\partial F_{03}}{\partial t} - \frac{\partial F_{13}}{\partial x} - \frac{\partial F_{23}}{\partial y} - \frac{1}{x} F_{13} = 0.
\label{ecuación para A2 y A3}
\end{equation}
Then, for $\nu=1$,
\begin{equation}
\frac{1}{\alpha^2 x^2} \frac{\partial F_{01}}{\partial t}-\frac{\partial F_{21}}{\partial y}-\frac{\partial F_{31}}{\partial z}=0.
\label{ecuación para A1}
\end{equation}
Finally, for $\nu=0$,
\begin{equation}
\begin{aligned}
-\frac{\partial F_{10}}{\partial x}-\frac{\partial F_{20}}{\partial y}-\frac{\partial F_{30}}{\partial z}+\frac{F_{10}}{x}=0.
\end{aligned}
\label{ec para A0}
\end{equation}

Thus, we obtain a system of four equations for the six independent components of $F^{\mu\nu}$. From this, we can derive the wave function for the remaining components of the vector potential, that is,  the scalar part and the longitudinal component. Next, we will present three possible cases to explicitly obtain solutions for $A^0$ and $A^1$.

\subsection{Case $k_y=k_z=0$ and $A_3=A_2=0$}

We begin with the simplest case where the field only has components $A_0$ and $A_1$, while assuming normal incidence. From (\ref{ecuación para A1}) we find that $F_{01}$ is time-independent with $\omega=0$. From (\ref{ec para A0}):
\begin{equation}
\begin{aligned}
-\frac{\partial F_{10}}{\partial x}+\frac{F_{10}}{x}=0.
\end{aligned}
\label{ecuación para A0}
\end{equation}

From this, we obtain $F_{10}(x) = C_1 x$,
where $C_1$ is constant in both $t$ and $x$. Applying the Lorentz gauge condition, we arrive at the system of equations:
\begin{equation}
\frac{\partial A_0}{\partial x } -\frac{\partial A_1}{\partial t }= C_1 x, \quad \quad \frac{1}{\alpha^2 x^2}\frac{\partial A_0}{\partial t }- \frac{\partial A_1}{\partial x } - \frac{A_1}{x}=0.
\end{equation}
We find the solutions:
\begin{equation}
A_1(x)=-\frac{D_1}{x},\quad A_0(x)=\frac{C_1x^2}{2}+C_2,
\end{equation}
where $D_1$, $C_1$, and $C_2$ are constants. Notably, in the inertial limit $\alpha \rightarrow 0$ with $x \rightarrow 1/\alpha$, we recover the expected result that the scalar electric field component becomes constant.

\subsection{General Case}

We now proceed to the most general case, where no approximations are made. We take the equations in (\ref{ecuación para A2 y A3}) and multiply them by $k_y$ and $k_z$, then subtract and add them so they can be expressed in terms of the modes $( \mathbf{k}_{\perp}\cdot\mathbf{A})=k_yA_2+k_zA_3$ and $\mathbf{i}\cdot(\mathbf{k}_{\perp}\times\mathbf{A})=k_yA_3-k_zA_2$. Finally, using the appropriate gauge condition, we arrive at

\begin{equation}
\begin{aligned}
& \left\{\frac{1}{x}\left(\frac{\partial}{\partial x}\left(x \frac{\partial}{\partial x}\right)\right)+\frac{\omega^2}{a^2 x^2}-k_{\perp}^2\right\} \mathbf{i} \cdot\left(\mathbf{k}_{\perp} \times \mathbf{A}\right)=0, \\ & \left\{\frac{1}{x}\left(\frac{\partial}{\partial x}\left(x \frac{\partial}{\partial x}\right)\right)+\frac{\omega^2}{a^2 x^2}-k_{\perp}^2\right\} i\left(\mathbf{k}_{\perp} \cdot \mathbf{A}\right)=0 \\
& \left\{\frac{1}{x}\left(\frac{\partial}{\partial x}\left(x \frac{\partial}{\partial x}\right)\right)+\frac{\omega^2}{a^2 x^2}-k_{\perp}^2\right\} A^0=-\frac{i \omega^2}{\alpha^2 x^3} A^1, \\ & \left\{\frac{\partial}{\partial x}\left(\frac{1}{x} \frac{\partial}{\partial x} x\right)+\frac{\omega^2}{\alpha^2 x^2}-k_{\perp}^2\right\} A^1=0.
\end{aligned}
\label{ecuación para A1, A0, times y cdot}
\end{equation}
Thus, we obtain Bessel equations for $A^1$, $i(\mathbf{k}{\perp}\cdot\mathbf{A})$, and $\mathbf{i}\cdot(\mathbf{k}_{\perp}\times\mathbf{A})$, whose solutions are again given by $H^{(1)}_\nu$ Hankel functions. Meanwhile, $A^0$ can be derived from the Lorentz gauge condition (with Rindler metric),
\begin{equation}
i\omega A^0 -\frac{1}{x}\frac{\partial}{\partial x}\left(xA^1\right)-i(\mathbf{k}_{\perp}\cdot\mathbf{A})=0.
\end{equation}
Here we emphasize an important point: in the inertial case, $A^0$ is simply a constant, whereas in this system it is not—it is instead expressed in terms of Hankel functions and their derivatives.

Once $A^\mu$ is known, we can derive the $\mathbf{E}$ and $\mathbf{B}$ fields. For this general case, one can verify that the orthogonality property of the fields is preserved:  $
\mathbf{E} \cdot \mathbf{B} =0$. Meanwhile, the Poynting vector, when $\mathbf{i}\cdot(\mathbf{k}_{\perp}\times\mathbf{A})=0$, becomes:
\begin{equation}
\mathbf{S}=\mathbf{E} \times \mathbf{B} = 
\begin{pmatrix}
\frac{-i\omega}{\alpha^2x^2} \left( A^2 \frac{\partial A^2}{\partial x} + A^3 \frac{\partial A^3}{\partial x} \right), \;
-\frac{\partial A^0}{\partial x}\frac{\partial A^2}{\partial x}, \;
-\frac{\partial A^0}{\partial x}\frac{\partial A^3}{\partial x}
\end{pmatrix}.
\end{equation}
Unlike the inertial case, here the Poynting vector exhibits position dependence. However, it can be verified that by taking the inertial limit $\alpha \rightarrow 0$ with $x \rightarrow 1/\alpha$, $A^0$ becomes constant and we recover $k_x$ as the photon momentum perpendicular to the mirror.

\subsection{Degenerate Case}

In this section, we consider time-independent fields ($\omega=0$). This modifies the differential equations in (\ref{ecuación para A1, A0, times y cdot}), where the frequency-dependent potential has now been eliminated, 
\begin{equation}
\begin{aligned}
\left\{\frac{\partial}{\partial x}\left(\frac{1}{x} \frac{\partial}{\partial x}\left(x \right)\right)-k_{\perp}^2 \right\} A^1 & =0, & \left\{\frac{1}{x} \frac{\partial}{\partial x}\left(x \frac{\partial}{\partial x} \right)-k_{\perp}^2 \right\}\mathbf{i}\cdot(\mathbf{k}_{\perp} \times \mathbf{A}) & =0, \\
\left\{\frac{1}{x} \frac{\partial}{\partial x}\left(x \frac{\partial}{\partial x} \right)-k_{\perp}^2 \right\} A^0 & =0, & \left\{\frac{1}{x} \frac{\partial}{\partial x}\left(x \frac{\partial}{\partial x} \right)-k_{\perp}^2 \right\} i(\mathbf{k}_{\perp} \cdot \mathbf{A}) & =0.
\label{ecuaciones caso degenerado}
\end{aligned}
\end{equation}
The Lorentz gauge condition then becomes
\begin{equation}
\frac{1}{x}\frac{\partial A^1}{\partial x}+i(\mathbf{k}_{\perp}\cdot\mathbf{A})=0. 
\end{equation}
From (\ref{ecuaciones caso degenerado}), we observe that the electrostatic potential $A^0$ will be a function $H_0^{(1)}(ik_\perp x)$, and for the case $k_\perp=0$, it becomes:
\begin{equation}
A^0(x)=C_1 \ln |\alpha x|+C_2.
\label{vac}
\end{equation}
Again, we can calculate the corresponding magnetic and electric fields, and it is particularly interesting to observe that in this case, the dot product between them equals
\begin{equation}
\mathbf{E}\cdot\mathbf{B}=\frac{\partial}{\partial x}\left[ \mathbf{i}\cdot\left( \mathbf{k}_{\perp}\times \mathbf{A}\right) A^0 \right] ,
\end{equation}
which is generally non-zero, when $\left( \vec{k}_{\perp}\times \vec{A}\right)\neq 0$. This constitutes a residual field. Due to the acceleration, we experience a fictitious force manifesting as an electrostatic field. These are virtual photons with an extra degree of freedom but a more physical interpretation can be given in what follows.

\subsection{Connection with Electro-vacuum solutions}

Although it may appear that these solutions are nonphysical, one must recall the existence of linear static fields in homogeneous Maxwell equations in flat space. These arise from the presence of charge distributions at infinity, e.g. a very large capacitor, typically distributed at a grounded surface that corresponds to an equipotential. Indeed, replacing a linear potential in a flat $F_{\mu\nu}$ and computing the divergence yields the sought vacuum solutions with flat equipotentials (planes normal to $x$ in this case). Now the connection between (\ref{vac}) and a deformed capacitor becomes apparent: In the limit $x \mapsto 1/\alpha$ where space is flat and no acceleration is felt, the replacement $x=\delta + 1/\alpha$, with small $\delta = x-1/\alpha$, and the expansion of $\log(1+\delta) \approx \delta$ yields, for $x>0$,

\begin{equation}
A^0(x) \approx C_1 (\alpha x-1) + C_2. 
\end{equation}
The energy density due to this field is constant, so for an infinite space-time, it is expected to store an infinite amount of energy. Interestingly, for the deformed case, such a density is no longer homogeneous and becomes singular at the wedge. It is possible to fix $C_1, C_2$ such that the field vanishes at a conducting surface $x=x_m$, i.e., the mirror.

\section{General properties of the anomalous wave operator}
\label{Mathematical properties of the anomalous wave operator}
Finally, we can explain the spectral anomaly in the framework of operator theory. Initially, we have square-integrable functions that do not quantize the spectrum and do not belong to the Sobolev space $W^{1,2}[0,\infty)$ (standard definitions in \cite{Sobolev, Sobolev2}) because the boundary condition is not yet established, and the singular point at the origin is included in the domain; the derivatives of the wave function $\sqrt{x}H^{(1)}_\nu(ik x)$ are not continuous at $x=0$. Therefore, the necessary conditions for the Sturm-Liouville theorem are not satisfied in $x \in [0,\infty)$. For applications in the detection of the Unruh effect, and in any physically relevant situation, one has to avoid the Rindler wedge. At such point, the FitzGerald contraction becomes singular. By placing the boundary condition at $x_m > 0$, all functions with a node at $x_m$ exhibit appropriate behavior $C_\infty$, and the potential $-1/x^2$ is regular. In this domain, the conditions of the Sturm-Liouville theorem are satisfied, and the functions belong to the Sobolev space $W^{m,p}[x_{\rm{m}},\infty)$, with $p = 2$ and $m\in \mathbb{N}$.

With the Sturm-Liouville theorem ensured, the operator (\ref{operador autoadjunto}) is self-adjoint, and our solutions form a complete set of orthonormal functions. Therefore, we obtain a resolution of the identity for all these functions in the form 
\begin{equation}
\begin{gathered}
\sum_n|\mathcal{N}_n|^2 H_\nu^{(1) *}\left(i k_n x\right) H_\nu^{(1)}\left(i k_n x^{\prime}\right) \sqrt{x x^{\prime}}=\delta\left(x-x^{\prime}\right), \\
\text { for } \quad x, x^{\prime} \in\left(x_m, \infty\right),
\end{gathered}
\end{equation}
where $k_n=k_{\perp n}=r_n(\nu)/x_m$, $\nu=i \omega/\alpha$ and $\mathcal{N}_n$ is the normalization factor that depends on the index. The relation is ensured by the orthogonality of the inner product:
\begin{equation}
\int_{x_m}^{\infty} d x|\mathcal{N}_n|^2 x H_\nu^{(1) *}\left(ik_{n} x\right) H_\nu^{(1)}\left(ik_{n^{\prime}} x\right)=\delta_{n n^{\prime}}.
\end{equation}

It is important to note that our treatment leads to convergent integrals by excluding $H^{(2)}_\nu$. This differs considerably from the result obtained in \cite{Berry, Crispino} for the modified Bessel functions $K_\nu$, since there is no clear distinction between functions that decay and grow exponentially. In fact, we have the identities \cite{abramowitz}:
\begin{equation}
K_\nu(z)= \begin{cases}\frac{1}{2} \pi \mathrm{i}^{v \pi \mathrm{i} / 2} H_\nu^{(1)}\left(z \mathrm{e}^{\pi \mathrm{i} / 2}\right), & -\pi \leq \mathrm{arg} z \leq \frac{1}{2} \pi \\ -\frac{1}{2} \pi \mathrm{i}^{-v \pi \mathrm{i} / 2} H_\nu^{(2)}\left(z \mathrm{e}^{-\pi \mathrm{i} / 2}\right), & -\frac{1}{2} \pi \leq \mathrm{arg} z \leq \pi\end{cases}
\label{modifiedbessel}
\end{equation}
where the left-hand side could be easily confused with a growing function. This distinction is crucial for future calculations of integrals in transition amplitudes for the Unruh effect. With these results, it will be possible to calculate the Klein-Gordon inner product between the solutions of the accelerated case  (Hankel function) and the solutions of the inertial case (sine functions). The overlap between such solutions, i.e $\phi_q$ and the accelerated case $\psi_{n, k_{\perp}}$, is
\begin{equation}
\left\langle\phi_q^*, \psi_{n, k_{\perp}}\right\rangle=\frac{e^{i\left(\omega_q-\omega_n\right) t}}{\alpha}\left[\frac{\omega_n-\omega_q}{(2 \pi)^{1/ 2}}\right]|\mathcal{N}_n|^2 \delta^{(2)}(\mathbf{k}_{\perp}-\mathbf{q}_{\perp})\int_{x_m}^{\infty} d x \sin \left[q\left(x-x_m\right)\right] \sqrt{x}H_{i\omega/\alpha}^{(1)}(i k_{\perp}x),
\label{overlap integral}
\end{equation}
where the dispersion relations are $\omega_q = \sqrt{q^2+q_{\parallel}^2}$ and $\omega_n = \omega(k_{\perp},k_{\parallel})$, with $k_{\perp}=k_n$ as in (\ref{cuantización de H}). 
The inner product is indispensable in the computation of the Bogoliubov transformation under the canonical formalism of quantization. The observable $\hat{A}$ in an inertial frame is transformed into an observable $\hat{B}$ in a non-inertial frame by a linear superposition. The results will allow to compute the corresponding coefficients and obtain the number expectation value $\langle N \rangle$, \textit{in vacuo}, related to the phenomenon of particle production. One should note here that a single-mode theory already shows interesting results in a second-quantized regime, as computed in \cite{miguel}. However, adding over all possible modes involves the more challenging convergence of overlaps as in (\ref{overlap integral}), but the convergence is now ensured.

\section{Physical effects of emergent modes and inner products}
\label{sec:physical}

\begin{figure}[h] 
    \centering 
    \includegraphics[width=1.0\textwidth]{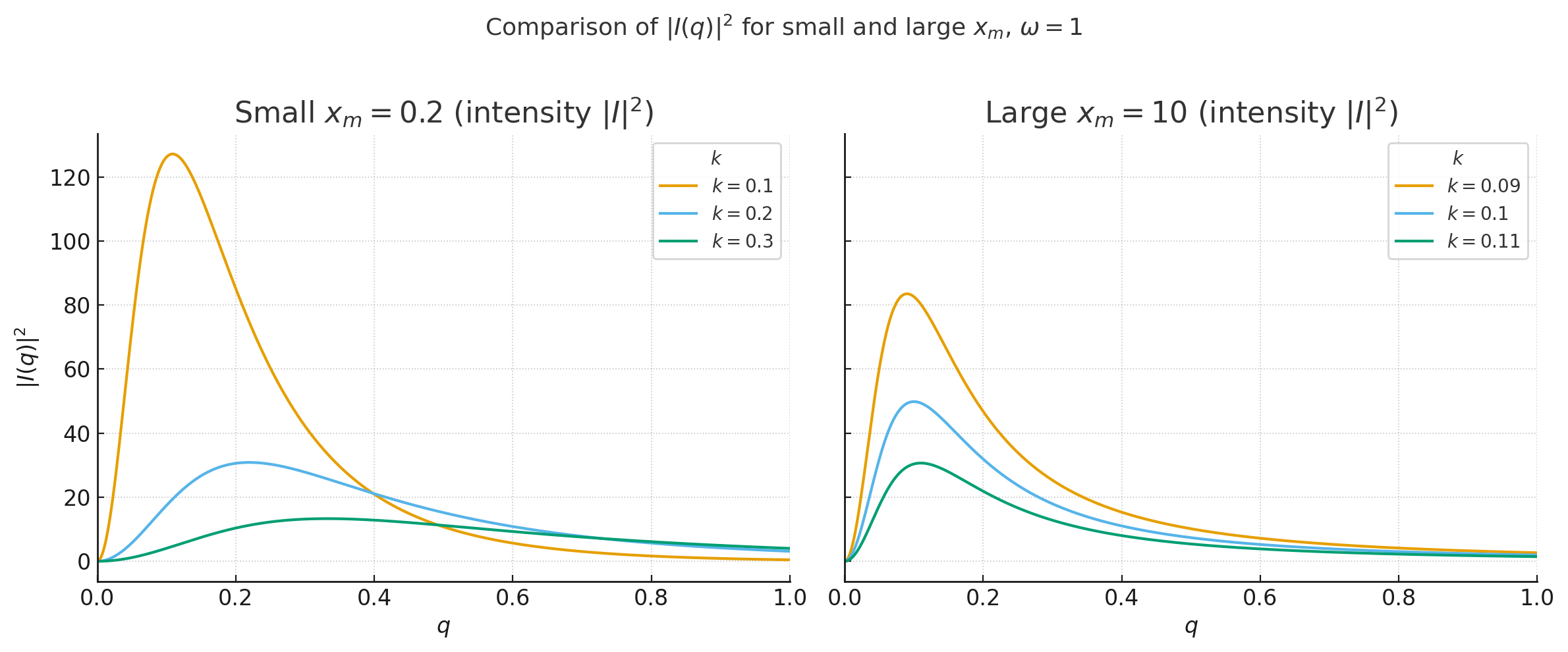}
    \caption{Intensity $|I|^2$ in two regimes. Left panel: mirror near singularity, equivalent to large accelerations; the curves display maxima at $q=k_n$ and are more dominant for low energies. Right panel: similar behavior for large mirror position into the Minkowski limit. 
} 
    \label{I1}
\end{figure}

\begin{figure}[h] 
    \centering 
    \includegraphics[width=0.6\textwidth]{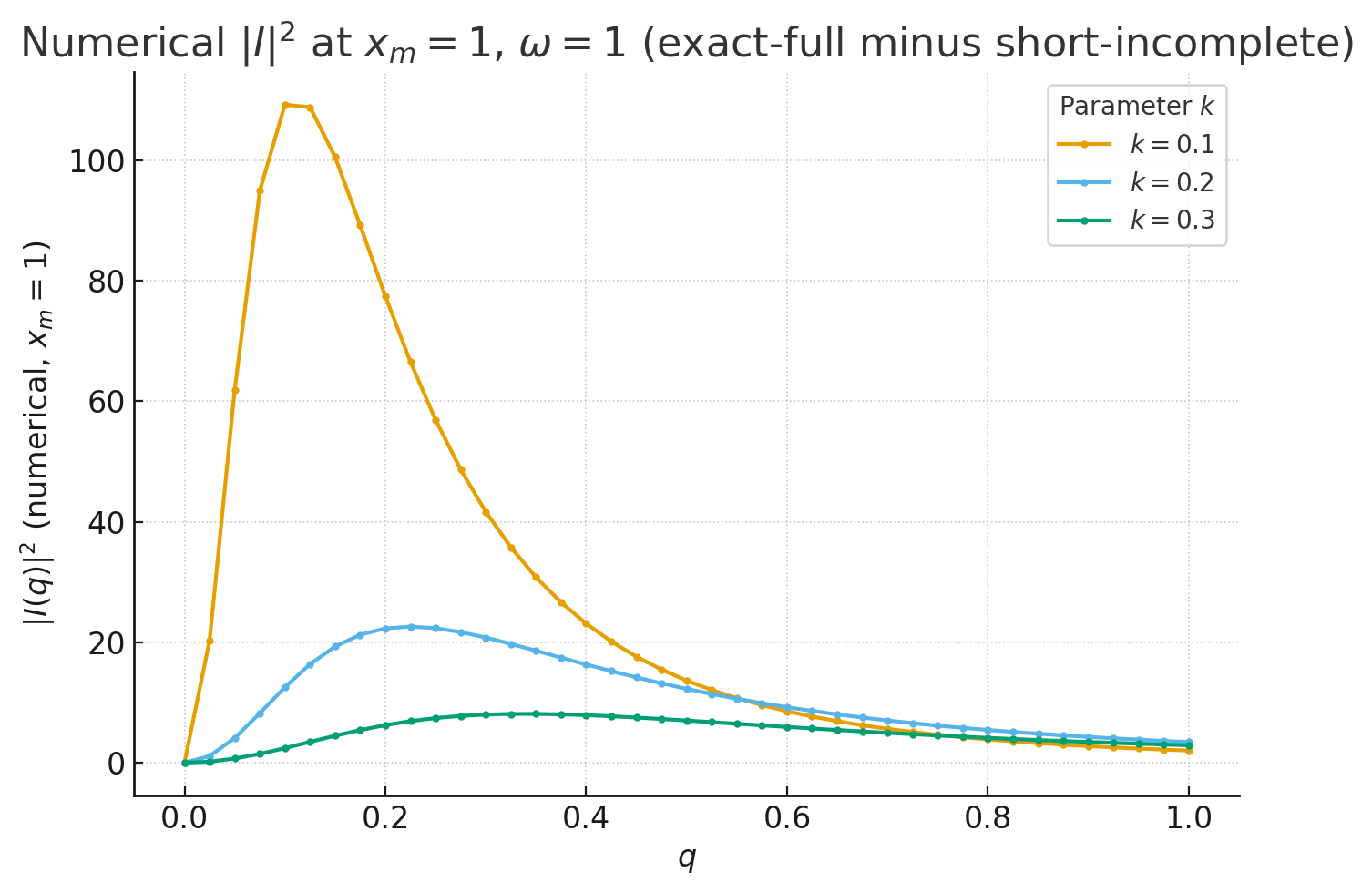}
    \caption{A numerical evaluation of $|I|^2$. The physical interpretation remains valid in this intermediate regime $x_m \approx 1$. The integral is divided into regions that include a full domain $0<x<\infty$ minus an incomplete part $0<x<x_m$ evaluated by adaptive quadrature.
} 
    \label{I2}
\end{figure}

The inner product (\ref{overlap integral}) between accelerated and static solutions gives an account of how an observer --holding the mirror initially at rest -- suddenly moves with acceleration $\alpha$ and interacts with the new modes; the latter are discretized. Although it is well known that all frequencies are affected by the acceleration, in our case it is possible to interact only with those that satisfy the boundary condition, and therefore, all calculations related to total radiated power (classical) and particle number (quantum-mechanical) detected by the mirror's motion pertain to discrete energies and the Hilbert spaces alluded in the previous sections. This puts a natural IR cutoff as the longest possible wavelength inside the $-1/x^2$ potential with boundary condition at $x_m$. The UV behavior will be established by asymptotic estimates; apart from transverse momentum conservation and Bohr frequencies in exponents, the formula (\ref{overlap integral}) contains the following relevant integral:

\begin{equation}
I \equiv \int_{x_m}^{\infty} \sin[q(x-x_m)] \sqrt{x} H_{i\omega/\alpha}^{(1)}(ikx)dx.
\label{relevantI}
\end{equation}
The result can be given in terms of Meijer-G functions, but numerical and analytical approximations must be used in order to extract physical information. The mirror's position relative to the wedge gives a hint on its decaying behavior using the following expressions:
\begin{equation}
I_{x_m \to \infty} \;\approx\;
\frac{2}{\pi i}\, e^{\pi \omega / 2\alpha}\,
\sqrt{\frac{\pi}{2k}}\,
\frac{e^{-k x_m}}{\sqrt{x_m}}\;
\frac{q}{k^{2} + q^{2}}\,,
\end{equation}
for flat regions, and
\begin{equation}
I_{x_m \to 0} \;\approx\;
\frac{e^{\pi \omega / 2\alpha}}{i \sqrt{k^{2} + q^{2}}}
\left[
\sin(q x_m)\,
\frac{\cos(\tfrac{\omega \eta}{\alpha})}{\cosh\!\left(\tfrac{\pi \omega}{2\alpha}\right)}
-\;
\cos(q x_m)\,
\frac{\sin(\tfrac{\omega \eta}{\alpha})}{\sinh\!\left(\tfrac{\pi \omega}{2\alpha}\right)}
\right],
\quad
\eta = \operatorname{arcsinh}\!\left(\frac{q}{k}\right),
\end{equation}
close to the singularity. Recalling the Bogoliubov transformation between inertial modes $A$ and non-inertial modes $B$

\begin{equation}
\hat{A}_q = \sum_{n}
\left[
\left( \frac{ \langle \phi_q , \psi_n \rangle^* }{ 2 \sqrt{ \omega_q \omega_n } } \right)
\hat{B}_n
+
\left( \frac{ \langle \phi_q^*, \psi_n \rangle }{ 2 \sqrt{ \omega_q \omega_n } } \right)
\hat{B}_n^{\dagger}
\right],
\end{equation}
one finds that (\ref{relevantI}) controls the weight per mode, and the sum of squares controls the average. A transition amplitude can be obtained via $|_B\langle \nu-1|A_q| \nu \rangle_B|^2$, for a field operator $A$ in the non-inertial basis $|\cdot\rangle_B$ with occupation number $\nu$. This quantity can be shown to be proportional to $|I|^2$, which is shown in Fig. \ref{I1} and \ref{I2}. In addition, the average $\langle A^{\dagger}_q A_q \rangle_{vac}$ can be obtained by adding all modes $\sum_n |I(q,k_n)|^2$, which has the appropriate exponential decay in $k_n$.

In the limit $\alpha \to 0, x_m \to 1/\alpha$, the transitions coincide. However, for other sets of parameters, the transitions offer a $q$-dependent distribution resembling Fano or Lorentzian shapes, depending on the mirror's position. It is not surprising to see that the photon momentum $q$ resonates with a characteristic $k=k_n$, making the weights more prominent for such values. The widths also increase with $k$ as their amplitude diminishes. Indeed, the weight vanishes exponentially (for large $x_m$) or quadratically (for small $x_m$) with increasing mode $k$, which corresponds to a very low probability of finding higher energy modes for particle production. This analysis is entirely new in the subject of Unruh radiation, thanks to our mathematical treatment.

\section{Conclusions}
\label{Conclusions}

In this paper we found spectrum quantization when \textit{only one} obstacle or boundary condition accelerates towards empty space; this was done for relativistic and non-relativistic equations. In both scenarios, the potential that allows quantization has the shape of a well complemented by the wall. It should be noted that in both cases, quantization occurs only in the region along the direction of uniformly accelerated motion, while behind the boundary condition, there is a continuous spectrum of energies.

For the relativistic case, the Klein-Gordon equation already displays the anomalous potential $-1/x^2$, where appropriate solutions must be selected. Maxwell's field equations also exhibit this property, and with a conceptual experimental setup, we can reduce the complexity of the equations to obtain the explicit solution for the field. It is worth noting that the resulting spectrum due to this quantization condition, to our knowledge, is reported here for the first time. The solutions to the problem were found for all values of $\alpha$, including large and small regimes. For instance, in the neighborhood $ 0 \leq \alpha < \varepsilon,$ with $\varepsilon \ll 1$, the wave functions and energies are non-analytic, indicating that a perturbative power expansion would not be possible, yet the limit $\alpha \rightarrow 0$ can be taken safely. This has important implications in the use of Feynman diagrammatics for transition amplitudes in the Unruh effect.

We find significant differences in the physical fields between the inertial and uniformly accelerated cases. First, the scalar component $A^0$, which was previously a constant, now satisfies a Bessel-type equation. In the inertial case, when taking zero frequency, $\mathbf{E}$ and $\mathbf{B}$ automatically vanish due to the dispersion relation. However, in the degenerate case with $\omega=0$, we obtain non-zero electric and magnetic fields, which additionally contain virtual photons with an extra polarization mode. Yet another interpretation was given in connection with electrovac solutions. This consequence might show in the \textit{little group} of the particle in Rindler space: The Christoffel symbols generate an additional term that plays the role of an effective mass in the differential equations, due to the fictitious force induced by acceleration. Consequently, when the momentum and Pauli-Lubanski vector are no longer light-like, the helicity state is no longer restricted to $\pm 1$, allowing the inclusion of intermediate states. These results motivate future study of how the \textit{little group} should be modified in this context, and the identification of new invariants to properly characterize particles in uniformly accelerated motion.

We would like to emphasize the importance of correctly defining the physical solution to the problem by choosing carefully the domain of the operators, which was done by picking one Hankel function $H^{(1)}$. This also involves the incorporation of the Jacobian in Rindler's transformation and the boundary conditions at infinity, ensuring thus the validity of the Sturm-Liouville theory of self-adjoint operators. Having a well-defined complete set of quantized solutions will lead to the correct computation of overlap integrals, i.e. (\ref{overlap integral}), as the convergence is now ensured. This is a key point in a more rigorous treatment of the wave functions and their role in the Unruh effect. The behavior and fast decrease of the weight distributions as a function of photon's momentum was provided.

\section*{Acknowledgments}
Financial support from VIEP project 100518931
BUAP-CA-289 is acknowledged.

\bibliographystyle{plainnat}

\providecommand{\noopsort}[1]{}\providecommand{\singleletter}[1]{#1}

\end{document}